\documentclass{aa} 
\usepackage{txfonts}
\usepackage{multirow}
\usepackage{graphicx,epsf}
\usepackage{url}
\usepackage{amsmath,graphicx,adjustbox,setspace, sidecap, float}
\usepackage[toc,page]{appendix}
\usepackage{gensymb}
\usepackage{fancyhdr}
\usepackage{natbib}
\usepackage{caption,subcaption}
\usepackage{threeparttable}
\usepackage{color}

\usepackage[dvipsnames]{xcolor}

\newcommand{\sig}{\Sigma}
\newcommand{\phip}{\phi_{\rm p}}
\newcommand{\Mp}{M_{\rm p}}

\newcommand{\rp}{r_{\rm p}}
\newcommand{\Ms}{M_{\star}}
\newcommand{\Me}{M_{\oplus}}
\newcommand{\Msun}{M_{\odot}}
\newcommand{\OmegaK}{\Omega_{\rm K}}

\newcommand{\Fg}[1]{Figure~\ref{fig:#1}}
\newcommand{\Tb}[1]{Table~\ref{tab:#1}}

\newcommand{\eq}[1]{Eq.~(\ref{eq:#1})\xspace}
\newcommand{\Eq}[1]{Equation~(\ref{eq:#1})\xspace}
\newcommand{\Eqs}[2]{Eqs.\ (\ref{eq:#1}) and (\ref{eq:#2})}

\newcommand{\se}[1]{Sect.~\ref{sec:#1}}
\newcommand{\Se}[1]{Section ~\ref{sec:#1}}

\usepackage[backref, pagebackref=false, hyperindex=true, breaklinks=true, colorlinks=true, urlcolor=magenta, linkcolor=magenta, citecolor=NavyBlue, citebordercolor={0 1 0}, pagecolor=red, bookmarks=true, filecolor=blue,bookmarksopen=true, plainpages=false, pdfpagemode=UseThumbs]{hyperref}

\begin{document}

\title{Hydrodynamical Simulations of Planet Rebound Migration in Photo-evaporating Disks}

\author{Beibei Liu\inst{1,2 \star},  Cl{\'e}ment Baruteau\inst{3}, Zhaohuan Zhu\inst{4}, Ya-Ping Li\inst{5} \and Sijme-Jan Paardekooper \inst{6}}

 \institute{Institute for Astronomy, School of Physics, Zhejiang University, Hangzhou 310027,  China \ 
 \ \email{bbliu@zju.edu.cn}
\and
  Center for Cosmology and Computational Astrophysics, Institute for Advanced Study in Physics, Zhejiang University, Hangzhou 310027,  China
\and
IRAP, Universit{\'e} de Toulouse, CNRS, CNES, Toulouse, France 
\and
Department of Physics and Astronomy, University of Nevada, Las Vegas, NV 89154-4002, USA
\and           
Shanghai Astronomical Observatory, Chinese Academy of Sciences, Shanghai 200030, China  
\and 
TU Delft, Faculty of Aerospace Engineering, Kluyverweg 1, 2629 HS Delft, The Netherlands
}

  \abstract
   {
   This study investigates the orbital migration of a planet located near the truncated edge of protoplanetary disks, induced by X-ray photo-evaporation originating from the central star. The combined effects of turbulent viscous accretion and stellar X-ray photo-evaporation give rise to the formation of a cavity in the central few astronomical units in disks. Once the cavity is formed, the outer disk experiences rapid mass loss and the cavity expands inside out. We have conducted 2D hydrodynamical simulations of planet-disk interaction for various planet masses and disk properties. Our simulations demonstrate that planets up to about Neptune masses experience a strong positive corotation torque along the cavity edge that leads to sustained outward migration -- a phenomenon previously termed {\it rebound} migration. Rebound migration is more favorable in disks with moderate stellar photo-evaporation rates of ${\sim}10^{-8} ~ \rm M_{\odot}\,yr^{-1}$. Saturn-mass planets only experience inward migration due to significant gas depletion in their co-orbital regions. In contrast, Jupiter-mass planets are found to undergo modest outward migration as they cause the residual disk to become eccentric. This work presents the first 2D hydrodynamical simulations that confirm the existence and viability of rebound outward migration during the inside-out clearing in protoplanetary disks.
   }
  
\keywords{Planet-disk interactions -- 
   Protoplanetary disks -- 
   Planets and satellites: formation -- 
   Methods: numerical }
    
\titlerunning{Rebound migration in photo-evaporating disks}
\authorrunning{Beibei Liu et al.}
\maketitle

\section{Introduction} 

Embedded planets tidally interact with their natal gaseous protoplanetary disks, inducing orbital migration \citep{Lin1993,Kley2012,Baruteau2013,Baruteau2014,Paardekooper2023}. Low-mass planets, like super-Earths and Neptune-mass planets, have modest influence on their disk's structure, with net angular momentum transfer originating from Lindblad and corotation resonances. The resulting wave torque and corotation torque (also known as horseshoe drag) have been extensively studied \citep{Goldreich1980,Ward1997,Tanaka2002,Baruteau2008,Paardekooper2010,Paardekooper2011,Benitez-Llambay2015,McNally2019,Wu2024}. On the other hand, more massive gas giant planets strongly perturb their disk and create annular gaps in the proximity of their orbits, leading to a complex coupling between their orbital motion and disk evolution \citep{Lin1986, Lin1996,Nelson2000,Lubow2006,Ida2013,Durmann2015,Dong2015,Kanagawa2018,Robert2018, Chen2020,Li2024}. Nonetheless, most of the literature studies have focused on planet-disk interaction in disks with originally smooth density profiles, leaving the understanding of how planets migrate in structured disks near the cavity-disk boundary relatively unexplored. 

\cite{Masset2006} first investigated how the torque balance is affected at disk locations where the gas density gradient is reversed, for instance near a cavity. For low-mass planets, the strength and direction of the corotation torque are sensitive to the local density gradient, and when the latter is sufficiently large and positive, planets can experience a large, positive corotation torque that effectively halts inward migration \citep{Masset2006, Hu2016}. \cite{Liu2017a} proposed that the inner disk is truncated at the magnetospheric cavity radius, typically located at orbital distances of $r{\lesssim}0.1$ au from the central star, depending on the topology and strength of the star's magnetic field \citep{Koenigl1991,Romanova2003,Zhu2024,Zhu2025}. It serves as a natural location for reversing the direction of planet migration. \cite{Liu2017a} showed via analytical calculations that, depending on the disk properties and planet mass, the corotation torque can become strong enough to overcome the negative, outer Lindblad torque. Such ideas have been further supported by dedicated hydrodynamical simulations \citep{Miranda2018,Romanova2019,Romanova2025}. In addition, as the gas disk depletes, the magnetospheric cavity expands, resulting in a cessation of inward planet migration and a subsequent outward movement. This mechanism, known as rebound, can cause planet pairs to move out of mean-motion resonances \citep{Liu2017a,Huang2022,Hansen2024}, which could explain the numerous non-resonant planet pairs detected by the Kepler mission \citep{Fabrycky2014,Liu2017b}.

Stellar photo-evaporation plays a crucial role in the evolution of protoplanetary disks \citep{Alexander2014,Ercolano2017}. During the early stages, the dominant mechanism of disk evolution is turbulent accretion \citep{Ercolano2017} and the influence of stellar photo-evaporation is minimal\footnote{While magnetized winds can also drive disk angular momentum transport \citep{Bai2016}, we emphasize that such wind-driven planet migration - being strongly dependent on poloidal magnetic flux \citep{Aoyama2023} - represents a different regime from what we focus on in this work.}.  However, once the disk accretion rate falls below the stellar photo-evaporation mass-loss rate, the dissipation of the disk is governed by the photo-evaporative wind originating from the central star's high-energy radiation \citep{Alexander2006,Owen2011,Wang2017,Picogna2019,Picogna2021}. As a result, an inner disk cavity is formed, typically located at a few au. The gas inside this radius is rapidly accreted onto the central star, while the gas beyond it evaporates as a wind. Consequently, the inner disk cavity expands from the inside out. Building upon the concept of the rebound mechanism operated at the magnetospheric cavity radius,  \cite{Liu2022} have recently proposed that the large-scale retreat of the photo-evaporation-induced disk edge can naturally explain the orbital structure of the four giant planets in the Solar System. Besides, rebound-triggered dynamical instabilities may have broad implications, and could play a role in the eccentricity distribution of warm/cold Jupiters (see also \citealp{Debras2021} for a different mechanism).
 
In light of the pioneering study of \cite{Liu2022}, in this paper we conduct dedicated hydrodynamical simulations, for the first time, to verify the robustness of the rebound mechanism under various realistic conditions. We investigate the migration of a single planet with various masses, embedded near the boundary of the cavity in an inner truncated disk induced by stellar photo-evaporation. 

The paper is organized as follows. The model setup and numerical procedure are introduced in Section \ref{sec:method}. The migration of a single planet with four characteristic masses for the fiducial run is described in Section \ref{sec:single}. Section \ref{sec:parameter} explores the influence of the initial disk surface density profile, photo-evaporation mass-loss rate and disk aspect ratio on migration pattern. Finally, we summarize our key findings in Section \ref{sec:conclusion}.

\section{Method}
\label{sec:method}

\begin{table*}[]
\caption{List of code units and corresponding physical values.}
\centering
\begin{tabular}{ccccccc}
\noalign{\smallskip}
\hline\hline 
mass unit $M_0$ & length unit $r_0$ & time unit $T_0$ & time unit $P_0$ \\
\hline
$1~M_{\odot}$ & $10$ au & $\Omega_0^{-1}=(r_0^{3} /GM_{\odot})^{1/2}{\approx} 5$ yr & $2\pi/\Omega_0{\approx}31 $ yr \\
\hline
\end{tabular}
\label{tab:unit}
\end{table*}

\begin{table*}[]
\caption{Main parameters of the simulations.}
\label{tab:parameter}
\centering
\begin{threeparttable}
\begin{tabular}{ccccccccc}
\noalign{\smallskip}
\hline\hline 
Description  & $s=d\log\Sigma_0 / d\log r$  & $\dot{M}_{\rm PE} \ \rm [M_{\odot} \ yr^{-1}]$ & $h_0$ & planet [resolution]$^{a}$  & rebound status$^{b}$ \\ 
\hline
\centering \textit{run-fid} &  -0.5 &  $ 3.2 \times 10^{-8}$     &  0.052 & Neptune [M]   & Yes \\ 
   & &    & & super-Earth [H]   & Yes \\
      & &    & & Saturn [M]   & No \\
           & &    & & Jupiter [L]   & Yes \\

\hline
\textit{run-dl}    & -1 &  $ 3.2 \times 10^{-8}$  &0.052 & Neptune [M]   & Yes \\ 
\hline
\textit{run-ph}  & $-0.5$ &  $1.6 \times 10^{-7}$    & 0.052 & Neptune [M]   & No \\ 
\hline
\textit{run-pl} & $-0.5$ &  $ 6.4 \times 10^{-9}$  & 0.052 & Neptune [M]  & No \\ 
\hline
\textit{run-ah}  & $-0.5$ &  $ 3.2 \times 10^{-8}$  & 0.07 & Neptune [M]   & No \\ 
\hline
\textit{run-al}  & $-0.5$ &  $ 3.2 \times 10^{-8}$  & 0.04  & Neptune [M]   & Yes \\ 
\hline
\end{tabular}
\small 
a. L, M, H stand for the low, medium and high grid resolutions (see text). \\
b. Yes and No correspond to the magnitude of rebound migration being larger or smaller than $3\%$, respectively.
\end{threeparttable}
\end{table*}

We consider a razor-thin gaseous disk orbiting a solar-mass star and hosting one planet. A two-dimensional cylindrical coordinate system $(r, \phi)$ is adopted, with the frame centered at the stellar location. The disk is assumed to evolve through photo-evaporation due to the central star, radial turbulent transport modelled as a viscous accretion process, and disk-planet interactions. The governing equations are:
\begin{equation}
\begin{aligned}
& \frac{\partial{\sig} }{\partial{t}} + \nabla   \cdot  \left( \sig \vec{v}  \right) = \dot \Sigma_{\rm PE}, \\
& \frac{\partial{\vec{v}} }{\partial{t}} + \vec{v} \cdot \nabla \vec{v} = -\frac{\nabla P}{\sig} - \nabla \Phi + \vec{f}_{\nu},
 \label{eq:conservation}
\end{aligned}
\end{equation}
where $\Sigma$, $P$, $\vec{v}$ and $\vec{f}_{\nu}$ represent the gas surface density, pressure, velocity and viscous force per unit mass, all of which are integrated over the vertical extent of the disk.
The total gravitational potential is the sum of the direct contributions from the star and planet, and the indirect contribution of the planet related to the acceleration of the coordinate frame:
\begin{equation}
\begin{aligned}
 \Phi = &-\frac{G \Ms} {r} 
      -  \frac{G \Mp} {\left[r^2 - 2r \rp \cos(\phi-\phip) + \rp^2 + \epsilon^2 \right]^{1/2}}\\
      &+ \frac{G \Mp}{\rp^2} r \cos(\phi-\phip),
 \label{eq:poten_eq}
 \end{aligned}
\end{equation}
where $G$ is the gravitational constant, $\Ms$ is the stellar mass, and $\Mp$, $r_{\rm p}$ and $\phip$ are the mass, radial distance, and azimuthal angle of the planet. The softening parameter $\epsilon$ is set to $0.6$ times the disk's pressure scale height at the planet's orbital radius. We assume that the disk gas is described by a locally isothermal equation of state, where the gas temperature varies with radius but stays constant in time (the disk's aspect ratio is specified below). Besides, we neglect the disk's self-gravity and gas accretion onto the planet. Since the disk's self-gravity is discarded, so is the indirect term arising from the disk \citep{Crida2025}.

We take into account the effect of high-energy radiation from the central star on the disk gas. Intense stellar radiation heats the upper layers of the disk, causing the hot gas ${\gtrsim} 10^{3} \rm ~K$ to escape as a photo-evaporative wind \citep{Alexander2014,Ercolano2017}. This process, known as stellar photo-evaporation, is incorporated in our model by the source term $\dot \Sigma_{\rm PE}$ in the mass conservation equation (see Eq.~\ref{eq:conservation}). In this work, we focus specifically on stellar X-ray radiation. Since the X-ray luminosity of a solar-mass star remains relatively constant during the short timespan of the disk evolution that we consider (see \citealt{Kunitomo2021}), $\dot \Sigma_{\rm PE}$ is treated as time independent and only varies with $r$.

\cite{Owen2011,Owen2012} calculated a $\dot \Sigma_{\rm PE}(r)$ profile based on their radiation-hydrodynamical model. For a solar-mass star with an X-ray luminosity of $L_{\rm X}{=}10^{30} ~ \rm erg~s^{-1}$,  the total photo-evaporation rate is $\dot M_{\rm PE} {\approx} 1.6\times 10^{-8}  ~M_{\odot} ~ \rm yr^{-1}$ \citep{Owen2012}. Following the same approach as \cite{Bae2013}, we adopt the primordial disk prescription of \cite{Owen2012} and normalize $\dot \Sigma_{\rm PE}(r)$ to conserve the total integrated mass loss rate $\dot M_{\rm PE}{=} 2 \pi \int  r  \dot \Sigma_{\rm PE}(r) dr$ (see their Appendix B1). By default, we adopt \cite{Owen2012}'s standard mass-loss rate, but different $\dot M_{\rm PE}$ values are explored in \Se{PE}\footnote{We note that the outcomes depend on the photo-evaporation models. For instance, for the same stellar X-ray luminosity, \cite{Picogna2019}'s profile yields a higher $\dot \Sigma_{\rm PE}$ in the outer disk region and an overall higher integrated mass-loss rate. This leads to faster disk dispersal and planet outward migration less likely.}. 

To investigate planet migration in a photo-evaporating disk during the gas dispersal phase, we employ \href{https://github.com/charango/dustyfargoadsg}{\texttt{Dusty FARGO-ADSG}}, an upgraded version of the original $2$D Eulerian grid-based hydrodynamical \texttt{FARGO} code \citep{Masset2000}. We have implemented the calculation of $\dot \Sigma_{\rm PE}(r)$ as specified above. The inclusion of photo-evaporation implies that simulations are not scale free and require specifying code units. In our setup, the code's units of length, mass, and time are $r_0{=}10$ au, $M_0{=}M_{\star}{=}1 \Msun$, $ t_0 {=}  \Omega_{0}^{-1}{=} (r_{0}^3/G M_0)^{1/2}{\approx}5$ yr (see \Tb{unit}). In addition, we define $P_0 {=} 2\pi/\Omega_0$ and adopt it as the canonical time unit for the figures. 

We assume that the gas turbulent viscosity is given by $\nu {=} \alpha c_{\rm s} H$, with $\alpha{=}10^{-3}$, $c_{\rm s}$ the sound speed, $H {=} h\times r {=} c_{\rm s}/\Omega_{\rm K}$ the disk's pressure scale height, $h$ the disk aspect ratio,  and $\OmegaK{=}\sqrt{G M_{\star}/r^3}$ the Keplerian angular frequency. The initial profiles of the surface density and aspect ratio are given by $\sig_0(r) {=} \sig_0 (r/r_0)^s$ and $h(r) {=} h_0 (r/r_0)^k$, with default parameters $\sig_0=44 ~ \rm g\,cm^{-2}$, $s{=}{-}0.5$, $h_0{=}0.052$, and $k{=}0.25$. Our default disk temperature $T_{\rm d}$ is computed by assuming that the disk is heated up by stellar irradiation and cooled down by the black-body emission from two surface layers. This yields $ \theta  L_{\star}/(4 \pi r^2) {=} 2 \sigma_{\rm SB} T_{\rm d}^{4}$, where the grazing angle of starlight on the disk surface $\theta$ is set to $0.2$ rad, $L_{\star}{=}L_{\odot}$ is the luminosity of a solar-mass star, and $\sigma_{\rm SB}$ is the Stefan-Boltzmann constant.  Different values of $s$ and $h_0$ will be explored in Sections~\ref{sec:profile} and~\ref{sec:aspectratio}, respectively. A summary of the main parameters used in the simulations is provided in \Tb{parameter}.

The computational domain extends from $r_{\rm min} {=} 0.2 r_0$ to $r_{\rm max} {=} 10  r_0$ in the radial direction, and covers the full $2\pi$ in the azimuthal direction. Cells are sampled logarithmically in the radial direction and linearly in the azimuthal direction. Three grid resolutions are used: low ($N_{\rm r}{=}300$, $N_{\phi}{=}480$), medium ($N_{\rm r}{=}480$, $N_{\phi}{=}770$) and high ($N_{\rm r}{=}1200$, $N_{\phi}{=}1900$). The low resolution is used for Jupiter-mass planets ($318$ Earth masses), the medium resolution for the Neptune- and Saturn-mass planets ($17$ and $95$ Earth masses, respectively), and the high resolution for super-Earth planets ($3$ Earth masses). The values of $N_{\rm r}$  ensure that the full width of the planet’s horseshoe region (${\approx} 2.4r_{\rm p} \sqrt{q_{\rm p}/h}$, $q_{\rm p}$ is the planet-to-star mass ratio) is covered by about ten radial cells, thereby guaranteeing sufficient resolution of the corotation torque. 
The values of $N_{\phi}$ yield approximately square grid cells such that 
\begin{equation}
\begin{aligned}
N_{\rm r} \log\left(1 + \frac{2 \pi}{N_{\phi}}\right){\approx}\log \left(\frac{r_{\rm max}}{r_{\rm min}} \right).
 \label{eq:Nrphi}
 \end{aligned}
\end{equation}

Simulating disk evolution and planet migration over the disk lifetime remains numerically impractical in 2D. The disk slowly evolves viscously for most of its lifetime, and photo-evaporation becomes dominant only when its mass loss rate exceeds the viscous accretion rate, subsequently driving rapid disk dispersal \citep{Alexander2014}. We aim to investigate how planets migrate during this advanced phase of the disk evolution when it has opened an inner cavity. Thus, to optimize computational efficiency, we first evolve the disk in $1$D ($N_{\phi}{=1}$) without a planet, until the inner hole has just about formed. We then restart the simulations in 2D by including the planet. This second stage is hereafter referred to as the PE-dominated disk phase. In our $2$D simulations the planet is introduced at its current mass without considering any mass tapering treatment. Since the time for a giant planet to establish a gap is much shorter than both the simulation duration and the period required for rebound migration to occur (see \Se{single}), we do not expect that 
different planet mass release manners (slow vs. rapid increase) would affect our results.

In the $1$D simulations, a standard open boundary condition is applied at both the inner and outer radial edges of the grid. For the $2$D simulations, we employ wave-damping zones extending over $[0.2{-}0.35]r_0$ and $[8.2{-}10]r_0$ at the grid's inner and outer edge to avoid reflections of the planet wake. In this so-called wave-killing zone, the radial and azimuthal velocities are damped toward their initial profile, while the density is damped toward a threshold value of  $\Sigma{=}10^{-5} ~ \rm g/cm^2 $ -- the same as that used in the photo-evaporation module to reset the gas surface density wherever it falls below that value.

The specific disk torque onto the planet $\Gamma$ is given by 
\begin{equation}
\begin{aligned}
 \Gamma = \frac{1}{ M_{\rm p }} \int_{r }  \int_{\phi} \Sigma \frac{ \partial \Phi_{\rm p}}{\partial \phi} rd r d \phi = \frac{1}{ M_{\rm p }} \sum_r T(r),
\end{aligned}
\label{eq:Gamma}
\end{equation} 
with $\Phi_{\rm p}$ the planet's gravitational potential (second term on the right-hand side of Eq.~\ref{eq:poten_eq}), and the specific torque density $T(r)$ represents the $\phi$-averaged torque over the radial grid cells.

 \section{Single Planet Migration}
\label{sec:single}

In this section, we present the results of simulations with our fiducial model (\textit{run-fid} in \Tb{parameter}). We examine four planet masses: 3, 17, 95, and $318 ~\Me$, representative of super-Earth, Neptune, Saturn, and Jupiter-mass planets, respectively. All planets are placed at $r_0{=}10$ au at the beginning of the $2$D simulations, so that their orbit remains well outside the formation region of the photo-evaporative cavity.  The results for each planet mass are described in the following subsections.

 \subsection{Neptune-mass planet}
 \label{sec:Neptune}

\begin{figure}
    \centering
    \includegraphics[width=\hsize]{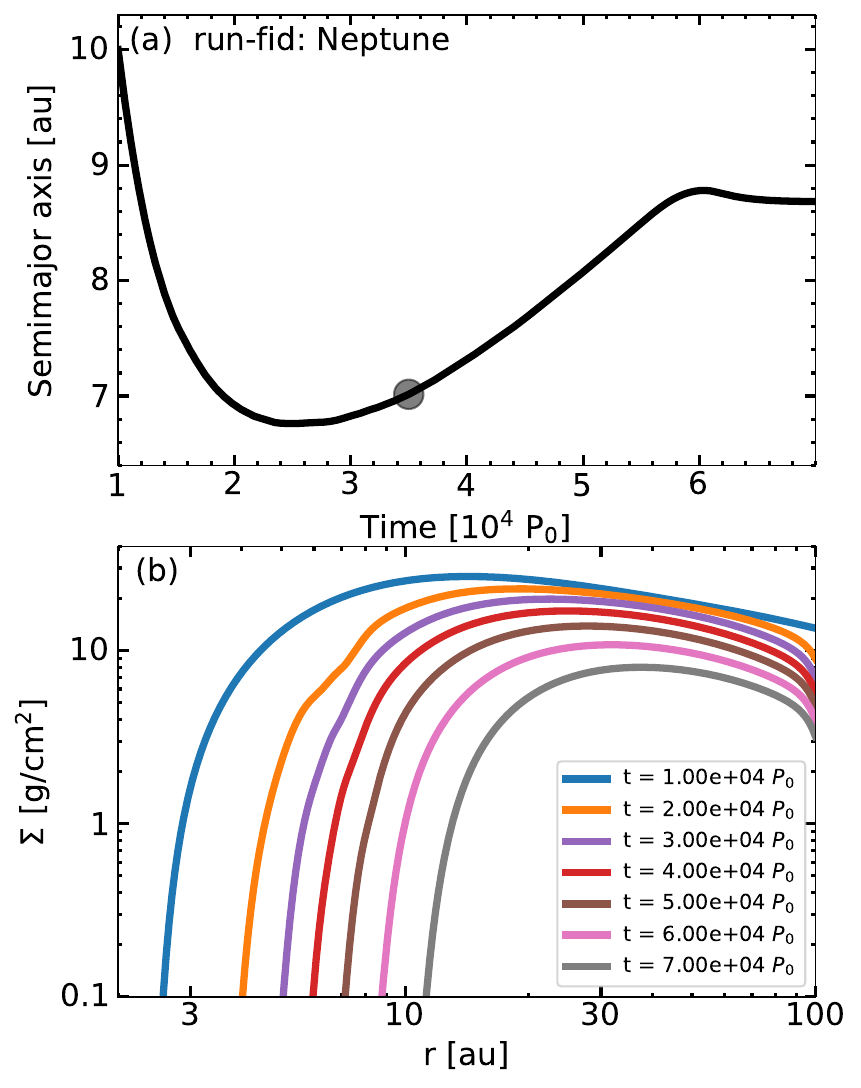}
    \caption{Migration of a Neptune-mass planet (panel a) and disk surface density evolution (panel b) in simulation \textit{run-fid}. In panel (a), the solid curve shows the time evolution of the planet's semi-major axis. The grey circle marks the planet position at the time shown in Figures~\ref{fig:sigN} and~\ref{fig:vertN}. The planet undergoes a reversal in migration direction (termed "rebound migration") as the inner photo-evaporative cavity expands inside-out. The video can be downloaded from: \protect\url{https://github.com/bbliu-astro/movies/blob/main/hydrodynamic_rebound/Neptune.gif}.}
    \label{fig:semiN}
\end{figure}

\begin{figure*}
    \centering
        \includegraphics[width=\hsize]{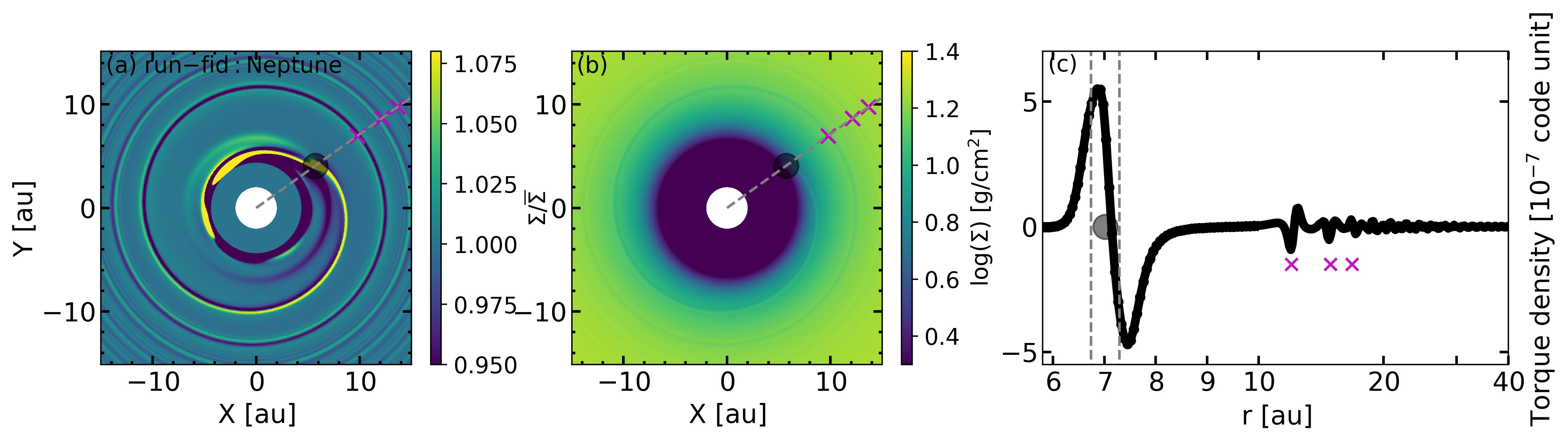}
    \caption{Surface density perturbation of $\Sigma/\overline{\Sigma}$ (panel a) and $\rm \log(\Sigma)$  (panel b), and specific torque density as a function of orbital distance (panel c) in \textit{run-fid} for a Neptune-mass planet at $t{=}3.5 \times 10^4~P_0$. The location of the planet is marked by a grey circle. In panel (b) data is averaged over $20$ snapshots in $1~P_0$, the purple crosses indicate the wiggle torque positions, and the width between the vertical dashed lines is $\Delta r_{\rm with} {=} 2 \max(r_{\rm H}, r_{\rm hs})$. 
    For visibility, the x-axis in panel c uses a linear scale up to 12 au and a logarithmic scale beyond that.
    } 
    \label{fig:sigN}
\end{figure*}

\begin{figure*}
    \includegraphics[width=0.48\hsize]{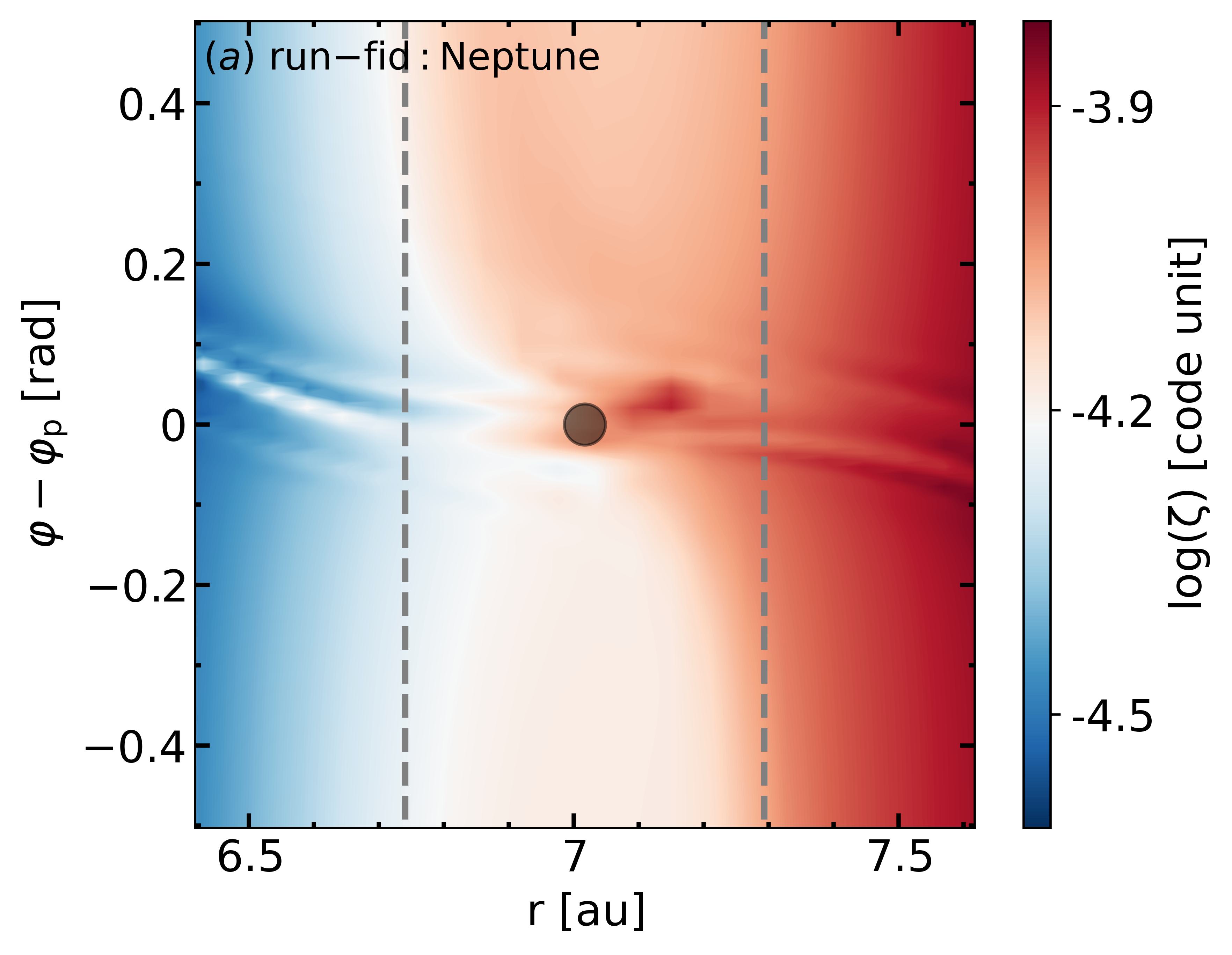}
    \includegraphics[width=0.5\hsize]{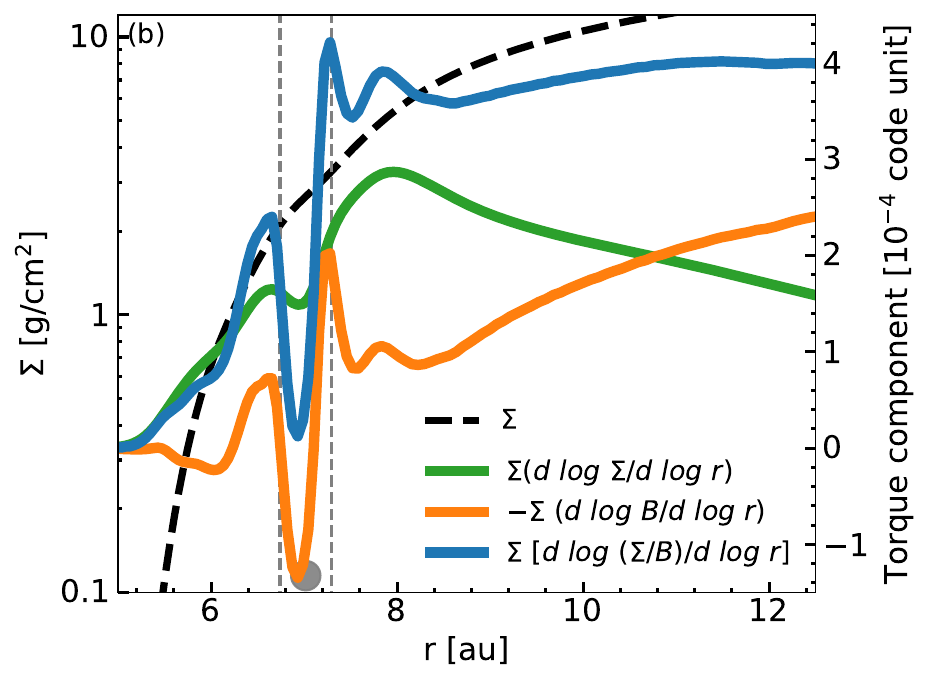} 
    \caption{
   Inverse vortensity ($\zeta$) close to the planet in the $r{-}\phi$ frame (panel a), surface density profile and corotation torque components (panel b) in \textit{run-fid} for a Neptune-mass planet at $t{=}3.5\times 10^{4} ~P_0$. In both panels, the width between the vertical dashed lines is $2 \max(r_{\rm H}, r_{\rm hs})$. The asymmetric $\zeta$ distribution between the upper and lower parts of the horseshoe region results in a strong, positive corotation torque.}
    \label{fig:vertN}
\end{figure*}

\Fg{semiN} illustrates the evolution of the semi-major axis and disk surface density for a Neptune-mass planet in the fiducial run (\textit{run-fid}). The planet is inserted at $t{=}10^{4} P_0$, after the inner hole has just formed. The planet first migrates inward, reaching ${\sim}6.7$ au at $t{=}2.5\times 10^{4}~P_0$ (\Fg{semiN}a). As the planet approaches the edge of the cavity, migration reverses and the planet moves outward over the next $3.5\times10^{4}~P_0$. This rebound migration coincides with the expansion of the inner photo-evaporative cavity (\Fg{semiN}b). The planet eventually stalls at ${\sim}8.7$ au after $6 \times 10^{4}~P_0$ as the disk has moved beyond the orbit of the planet.

\Fg{sigN} display the snapshot of the perturbed surface density in Cartesian coordinates at $t{=}3.5\times 10^{4}~P_0$. More specifically, the quantity $\Sigma / \overline\Sigma$ is displayed with $\overline\Sigma$, the instantaneous axisymmetric surface density in \Fg{sigN}a,  while the absolute surface density $\log(\Sigma)$ is shown \Fg{sigN}b. The inner hole is in the central $5.5$ au and the planet is at $x{=}5.7$ au, $y{=}4.1$ au (grey circle). The planet's spiral wakes are clearly visible, but the inner wake does not show up in the cavity since the disk's surface density there is reset to the minimum threshold value allowed in the simulation. The snapshot also highlights an asymmetric density perturbation in the planet's co-orbital region, with a positive (negative) density perturbation in front of (behind) the planet in the azimuthal direction. This suggests that a significant corotation torque is exerted on the planet. 
      
To gain further insight into the mechanism behind outward migration, \Fg{sigN}c shows the radial profile of the specific torque density $T$ given by \Eq{Gamma}. To minimize the effects of temporal fluctuations, $T$ has been averaged over $20$ snapshots within $1 ~P_0$. Not surprisingly, the gas interior to the planet exerts positive torques, while the gas exterior to the planet exerts negative torques. The total torque is a sum of these negative and positive components. The vertical dashed lines mark the region of the width around the planet $\Delta r_{\rm with} {=} 2 \max(r_{\rm H}, r_{\rm hs})$, where $r_{\rm H}{=} r_p\sqrt[3]{q_{\rm p}/3} $ is the planet Hill radius, $q_{\rm p}{=}M_{\rm p}/ M_{\star}$ and $r_{\rm hs}{=}1.2 r_{\rm p} \sqrt{q_{\rm p}/h} $ is the half-width of its horseshoe region \citep{Paardekooper2010}. This region is where the torque density is strongest. We point out that the two peaks in the torque density are not symmetric with respect to the planet, both in terms of radial position and amplitude. In particular, the inner peak in the torque density is larger than the outer peak, further indicating that the planet experiences a strong, positive corotation torque. We also note that beyond the location of the outer 1:2 Lindblad resonance, the torque density exhibits radial oscillations with negative and positive values and with a steadily decreasing amplitude as $r$ increases. This phenomenon, referred to as "wiggle torques" in \cite{Cimerman2023}, has been found in numerous hydrodynamical simulations \citep{Rafikov2012,Arzamasskiy2018,Dempsey2020}. It occurs where the wake crosses the $\phi{=}\phi_{\rm p}$ line as the spiral propagates (indicated by crosses in Figure \ref{fig:sigN}a and b). The net contribution of these oscillating torque components is very small compared to the torque components near the planet (within about $10\%$ of the planet's orbital radius).

 \Fg{vertN}a shows the inverse vortensity of the disk close to the planet's position. It is defined as
 \begin{equation}
  \zeta = \frac{ 2\sig}{ \left( \nabla \times \vec{v}  \right)_{\rm z}} = \frac{\Sigma}{B}, 
  \label{eq:vert}
 \end{equation}
 where  $B{=} (\nabla \times \vec{v})_{\rm z}/2$ is the second Oort constant related to the z-component of flow vorticity. In locally isothermal disks, the corotation torque ($\Gamma_{\rm c}$) features two components: one related to the local density gradient, and one to the local temperature gradient (see Eq. 49 in \citealp{Paardekooper2010}). Along the edge of the disk cavity, the former far exceeds the latter, so in practice $\Gamma_{\rm c} \propto \Sigma \times \  d \log \zeta /d \log r$ at the planet's orbital radius $r_{\rm p}$: the local radial gradient of $\zeta(r)$ determines the sign and strength of the corotation torque. \Fg{vertN}a highlights an azimuthal asymmetry in $\zeta$ within the planet's horseshoe region: the part in front of the planet in the azimuthal direction exhibits significantly higher $\zeta$ values than the part behind the planet, which is consistent with the density distribution in \Fg{sigN}a. This is further confirmed by the coloured curves in \Fg{vertN}b, which show that the $\zeta$ gradient across the planet is essentially set by the density gradient rather than by the vorticity gradient. The positive torque exerted by the horseshoe fluid elements flowing inward relative to the planet is thus much greater than the negative torque due to the horseshoe fluid elements flowing outward relative to the planet. This yields a net positive corotation torque on the planet, consistent with the observed outward migration.

The asymmetric corotation torque observed in our simulations is theoretically supported by \cite{Liu2017a}. Their analysis assumed rapid gas removal at a sharp disk edge, such that only the upper horseshoe stream contributes to angular momentum exchange, producing a maximal, positive corotation torque (termed ``one-sided torque''). Our realistic hydrodynamical simulations reveal a more moderate but consistent asymmetry, yielding a net positive torque that drives outward planet migration.
 
In summary, in the PE-dominated disk phase, a Neptune-mass planet can experience significant outward migration during the inside-out expansion of the photo-evaporative cavity. The net positive torque felt by the planet mainly arises from the asymmetric distribution of inverse vortensity in the planet's horseshoe region.

 \subsection{Super-Earth mass planet}
 \label{sec:super-Earth}

 \begin{figure*}
    \centering
    \includegraphics[width=0.33\hsize]{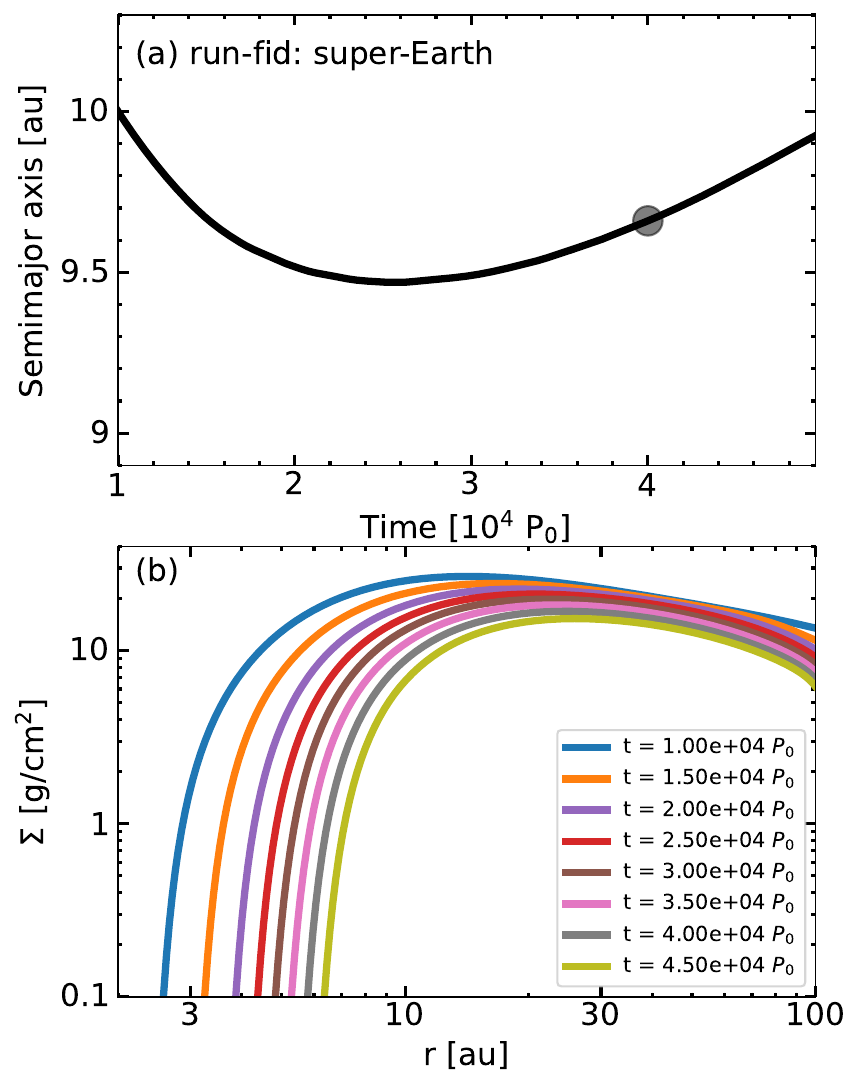}
    \includegraphics[width=0.33\hsize]{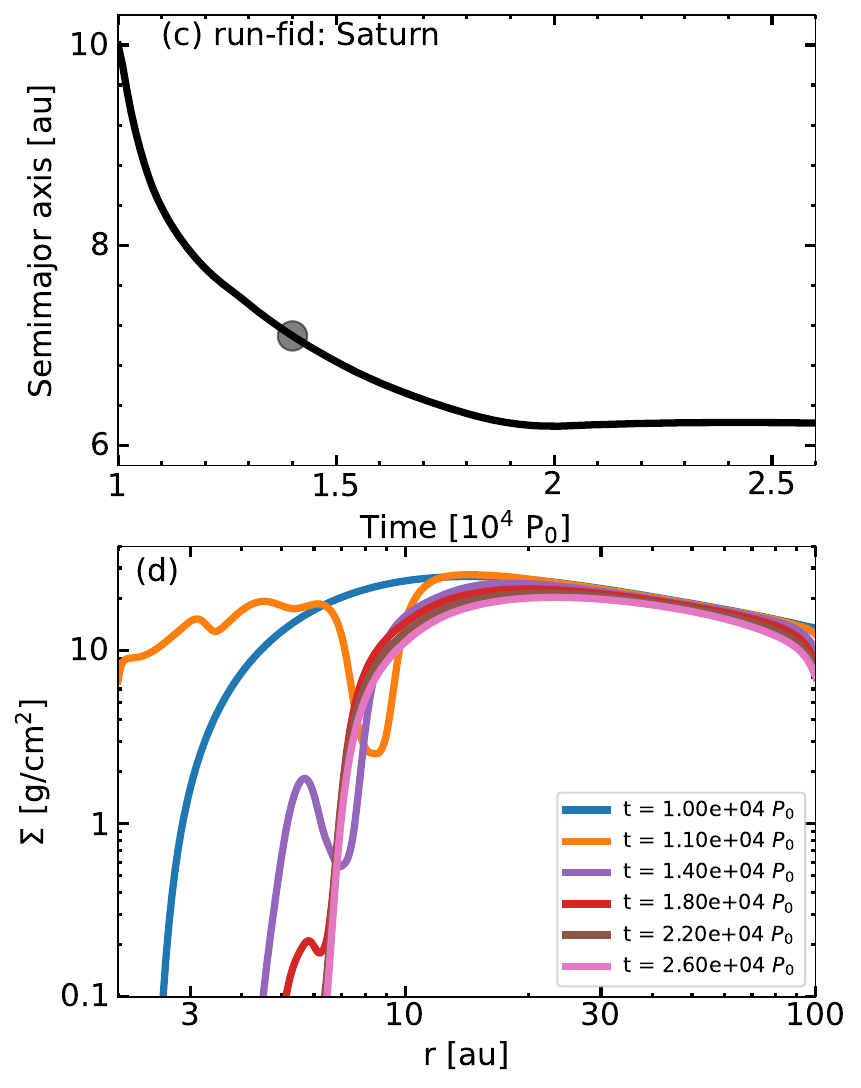}
    \includegraphics[width=0.33\hsize]{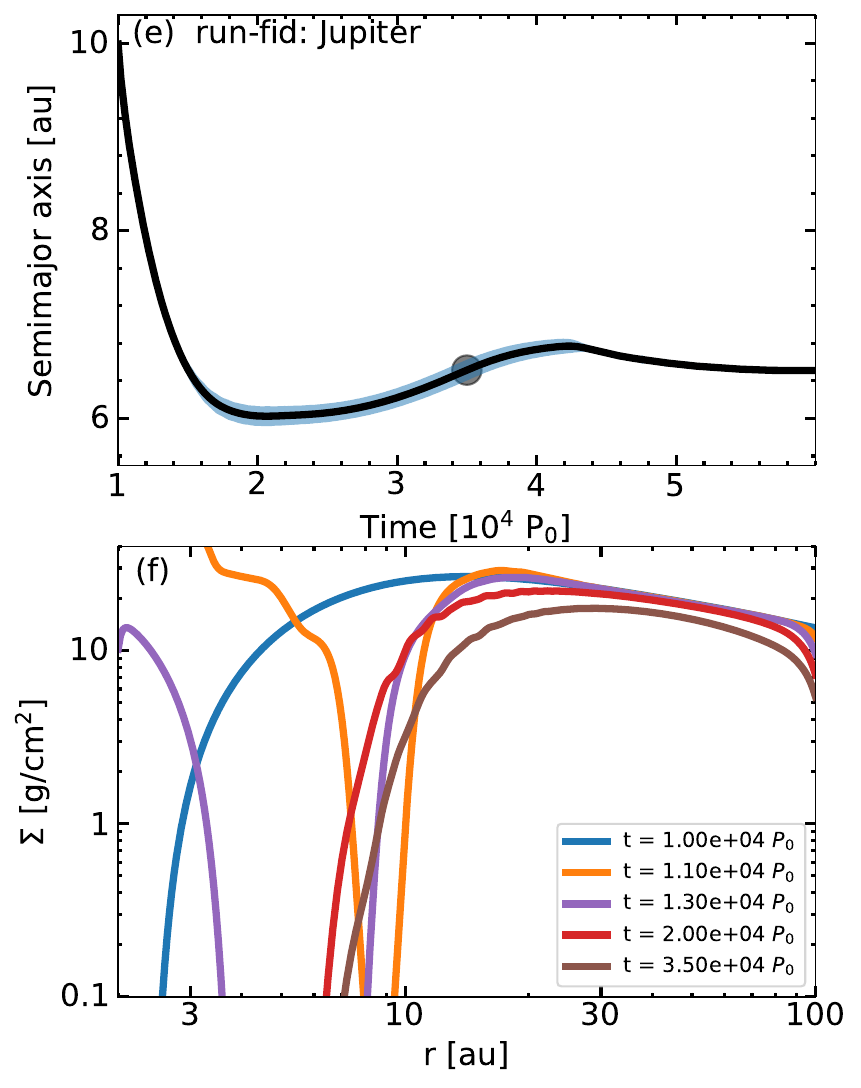}
    \caption{Evolution of semi-major axis (top) and disk surface density (bottom) in simulations \textit{run-fid} for a super-Earth planet of $3 ~M_{\oplus}$ (left), a Saturn-mass planet (middle) and a Jupiter-mass planet (right). In the upper panels the grey circles mark the planet's position used in \Fg{snapshot}, and the cyan area in the top-right panel shows the radial extent between the planet's pericentre and apocentre. The video can be downloaded from: \protect\url{https://github.com/bbliu-astro/movies/blob/main/hydrodynamic_rebound/superEarth.gif}.}
    \label{fig:semi}
\end{figure*}

\begin{figure*}
    \centering
     \includegraphics[width=\hsize]{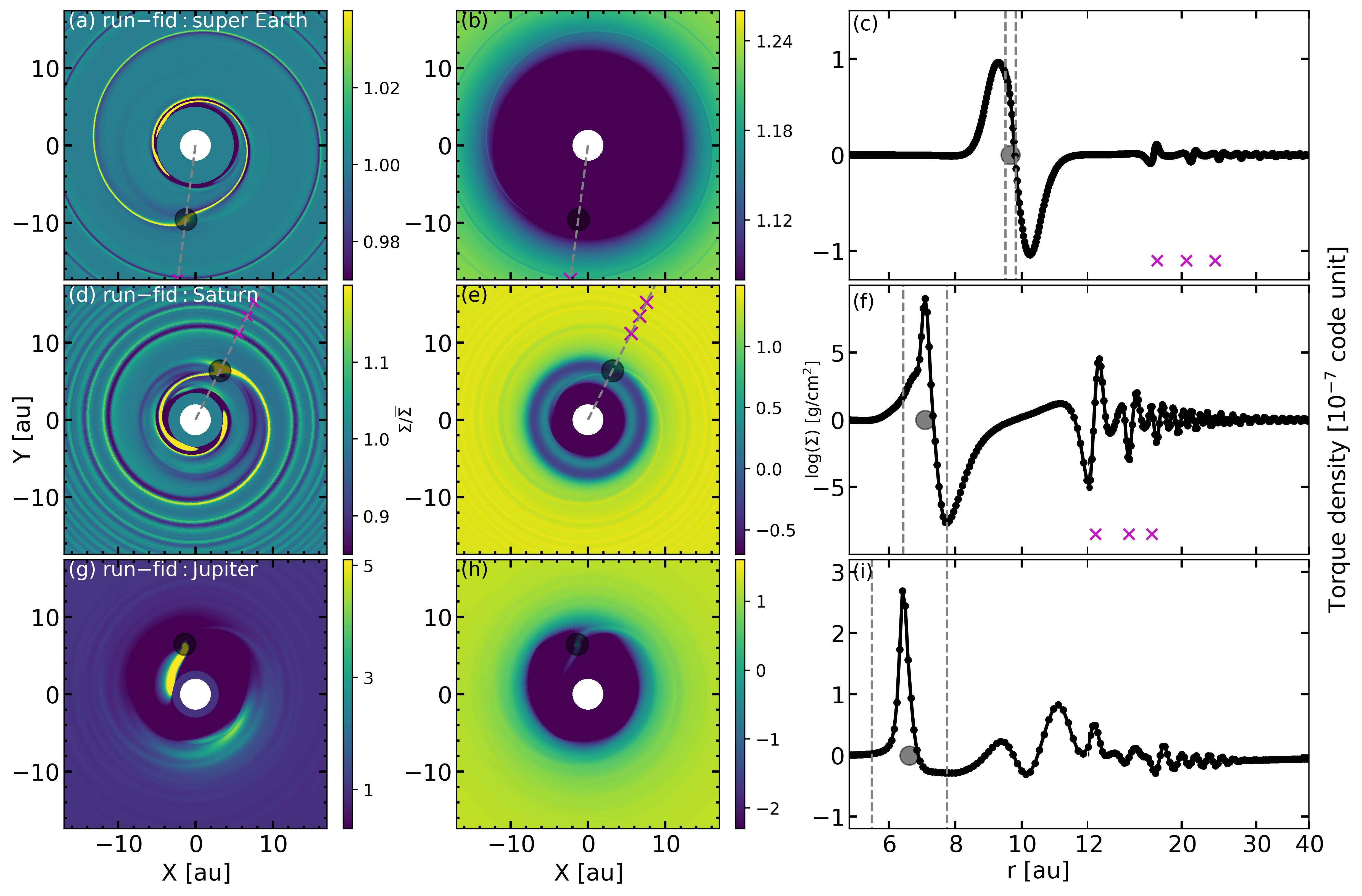}
    \caption{ 
   Surface density perturbation of $\Sigma/{\overline{\Sigma}}$ (left) and $\rm \log(\Sigma)$ (middle),  and specific torque density as a function of orbital distance (right) in \textit{run-fid} for a super-Earth planet of $3 \ M_{\oplus}$ (upper panels, $t{=}4 \times 10^{4}~P_0$), a Saturn-mass planet (middle panels, $t{=}1.4 \times 10^{4}~P_0$) and a Jupiter-mass planet (lower panels, $t{=}3.5 \times 10^{4}~P_0$). The location of the planet is marked by a grey circle. The data used to compute the torque density is averaged over $20$ snapshots in $1~P_0$, the purple crosses in the top and middle rows of panels indicate the wiggle torque positions, and the radial width between the vertical dashed lines is $\Delta r_{\rm with} {=} 2 \max(r_{\rm H}, r_{\rm hs})$. For visibility, the x-axes in the right panels use a linear scale up to 12 au and a in logarithmic scale beyond that.
   }
    \label{fig:snapshot}
\end{figure*}

We also performed a simulation with a super-Earth planet of $3~M_{\oplus}$. \Fg{semi}a shows that the planet's inward migration stalls at $r{\sim}9.5$ au near $t{=}2.5\times 10^{4}~P_0$, and then reverses as the cavity edge sweeps out. The planet goes back to ${\sim}10$ au at $t{=}5\times 10^{4}~P_0$ au\footnote{It is well possible that the planet would have further migrated outward if we had continued the simulation. However, after having consumed 100,000 core hours of computational resources, the planet evolution proceeded at an extremely slow rate and we manually terminated the simulation at $t{=}5\times 10^{4}~P_0$.}. 

Compared to the Neptune-mass planet run shown in the previous subsection, the density perturbation arising from the spiral wakes is weaker, at the $1{-}2\%$ level (\Fg{snapshot}a). The asymmetric density perturbation in the planet's horseshoe region is also reduced, though still clearly visible. Interestingly, we see that the torque density distribution differs a bit from the Neptune-mass planet case; in particular, the negative and positive peaks now have more equal amplitudes (\Fg{snapshot}c). Still, the torque distribution in the immediate planet vicinity (marked by the vertical dashed lines) reveals a substantial, positive corotation torque. In summary, our simulations show that the low-mass planets (both Neptune-like and super-Earth-like planets) exhibit similar rebound migration behavior during the expansion of the photo-evaporative cavity.

 \subsection{Saturn-mass planet}
 \label{sec:Saturn}

In the simulation with a Saturn-mass planet, we do not observe outward migration. The planet directly migrates into the disk cavity within $t{=}2\times 10^{4} ~ P_0$ (\Fg{semi}c). At $t{=}1.4 \times 10^{4} ~ P_0$, there is a considerable depletion of gas both inside the cavity and at the location of the gap carved by the planet (\Fg{snapshot}d \& e). In the gap, the disk exhibits a somewhat non-axisymmetric structure which we attribute to the onset of the Rossby wave instability (RWI). This and the (partial) gap formation around the planet could explain that the corotation torque is much less efficient at triggering outward migration than for the lower-mass planets investigated in the previous subsections.

 \subsection{Jupiter-mass planet}
 \label{sec:Saturn}

In contrast to the Saturn-mass planet, the Jupiter-mass planet did experience outward migration. Here the disk mass is too low for  Jupiter-mass planet to undergo type III migration, as this rapid migration typically requires higher disk masses and specific dynamical conditions (e.g., co-orbital mass deficit, \citealt{Masset2003,Peplinski2008}). 
We note that the planet opens a gap at a relatively fast pace. The torque onto the planet from the impulse approximation reads ${\sim}\Sigma r_{\rm p}^4 q_{\rm p}^2 \Omega_{\rm K}^2 (r_{p}/\Delta)^3$, where $\Delta{\simeq}H_{\rm p}$ \citep{Lin1993}. This gives the gap opening time approximately a few hundred orbits for a Jupiter-mass or Saturn-mass planet (\Fg{semi}d and f).  
The cavity take much longer time to expand from $2$ au to the approximate location of the planet. Thus, our setup describes a circumstance where the giant planet first opens a narrow gap. Later, stellar photoevaporation sets in and works in tandem to create a wide, common gap. 
In this case, the massive giant planet shortens the cavity opening time by reducing the flux across its orbit (also see \citealt{Rosotti2013}). 
We nevertheless do not expect qualitative differences in the outward migration pattern at different planet release times, although the detailed migration speed might differ. 

Our result is shown in \Fg{semi}e: after a rapid phase of inward migration, the planet migrates outward at a much slower pace from about 6 to 6.6 au. At $t{=}3.5 \times 10^{4} ~ P_0$, \Fg{snapshot}g and h shows a significant depletion of gas in the inner hole caused by both photo-evaporation and the planet wakes. In comparison to the previous simulations with lower planet masses, it is clear that the Jupiter-mass planet strongly perturbs its surrounding disk. In particular, the disk exhibits an asymmetric, crescent-like structure just outside of the cavity (see the density enhancement near 5 o'clock in \Fg{snapshot}g), and the gas in the cavity is clearly eccentric.

Our results share some analogy with those of \cite{Dempsey2021} and \cite{Li2021}. \cite{Dempsey2021} showed that in 2D simulations when the dimensionless quantity $\lambda{=} q_{\rm p}^2/(\alpha h^{3}){\gtrsim} 10{-}20$, migration is directed outward and the eccentricity of the disk beyond the planet goes from nearly circular to ${\gtrsim}0.2$. \cite{Li2023} pointed out that a different criterion may be needed to excite the disk eccentricity in 3D simulations.  We found $\lambda{\simeq}7$, despite that in our 2D simulation the planet is embedded in a cavity and therefore lacks an inner disk. In such a case, the disk can become eccentric and undergo apsidal precession at lower threshold values for $\lambda$ than found in \cite{Dempsey2021}. To further explore this, we computed the azimuthally averaged radial profile of the disk eccentricity following the method of \cite{Kley2006,Teyssandier2017}. Assuming that the gas behaves as a pressure-less particle and moves in a Keplerian orbit around the central star, we compute the eccentricity for each gas cell using its local position and velocity vectors. The eccentricity of the gas ring at the radial distance $r$ is then obtained by azimuthally averaging these cell values. The results are shown in \Fg{edisk} and compared with those of the simulations shown in Figures~\ref{fig:sigN} and~\ref{fig:snapshot}. We see that in the Jupiter-mass planet simulation, the eccentricity of the disk outside the cavity reaches about $0.2$, whereas for the lower-mass planets, it does not exceed $0.05$. The non-axisymmetric overdensity that forms outside the cavity could result from the RWI due to the strong density gradient across the cavity, or from a traffic jam due to the disk gas being eccentric \citep{Ataiee2013}. We note that in the simulation, the planet's eccentricity does not exceed $0.02$.

Furthermore, the torque density in this simulation is significantly altered and highly time-variable compared to the previous simulations with lower-mass planets. Instead of showing two lobes of similar shape and opposite signs, the torque density for the Jupiter-mass planet features a strong, positive contribution located close to the planet  (\Fg{snapshot}i). Since the gas is substantially depleted in the vicinity of the planet, this torque component is attributed to inner Lindblad torques within the eccentric disk, where the wake becomes largely bent near the inner disk edge. The torque density in the outer disk still exhibits oscillations that now extend to more distant separations as a result of the disk eccentricity. Finally, the planet migrates outward mainly because of the unbalanced Lindblad torques, where the inner positive torque largely dominates over the outer negative torque.

It is worth noting that \cite{Monsch2021} also explored the migration of a Jupiter-mass planet near the photoevaporating cavity during the disk dispersal phase. While their study confirmed that the planet can be halted at the disk edge, they did not observe the outward migration. We suspect this difference may originate from the distinct photoevaporation models employed.
\cite{Monsch2021} adopted the \cite{Picogna2019}'s photoevaporation profile. For a given stellar X-ray luminosity, this prescription produces a more significant mass loss in the outer disk regions and yields a higher total integrated mass-loss rate compared to the \cite{Owen2012}'s prescription used in our work. Consequently, the disk clears much faster in the \cite{Picogna2019}'s model, making the planet surfing on the expanding cavity more unlikely (see also \se{PE}).

\begin{figure}
    \centering
       \includegraphics[width=0.96\hsize]{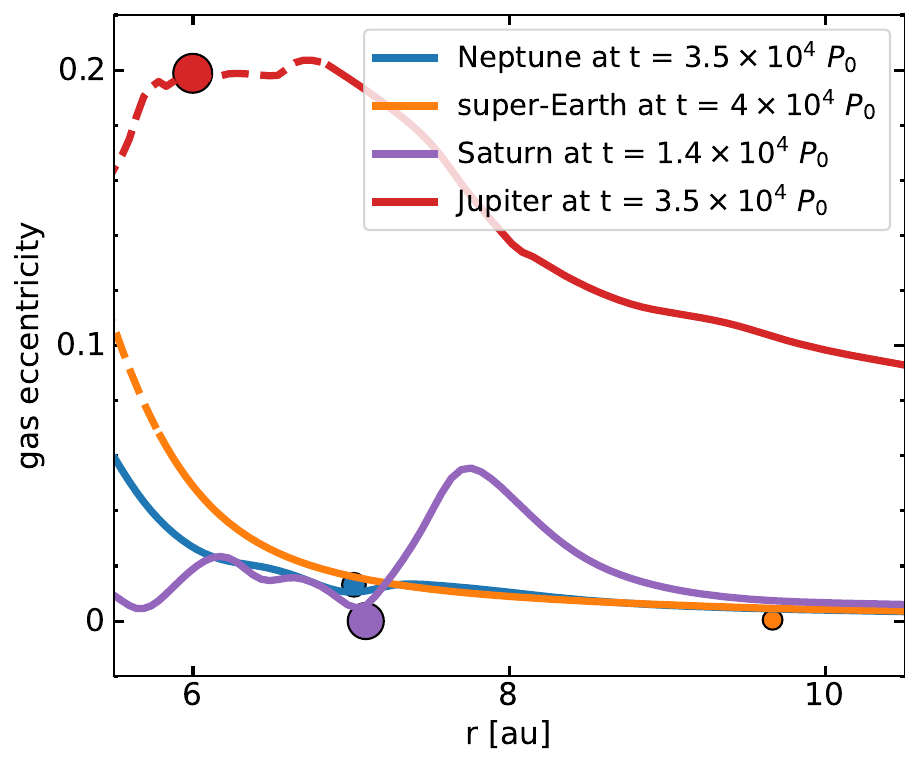}   
    \caption{Azimuthally-averaged radial distribution of the disk eccentricity at the same times as in Figures~\ref{fig:sigN} and~\ref{fig:snapshot}. The filled circles show the planet's location in each simulation. The dashed curves indicate the gas-depleted cavity region, where the calculation of the gas eccentricity is strongly affected by the reset in the gas density. The Jupiter-mass planet excites quite large eccentricity in its outer disk in contrast to the three less massive planets.} 
    \label{fig:edisk}
\end{figure}

 \section{Exploration of parameters space}
 \label{sec:parameter}

 \begin{figure*}
    \centering
       \includegraphics[width=\hsize]{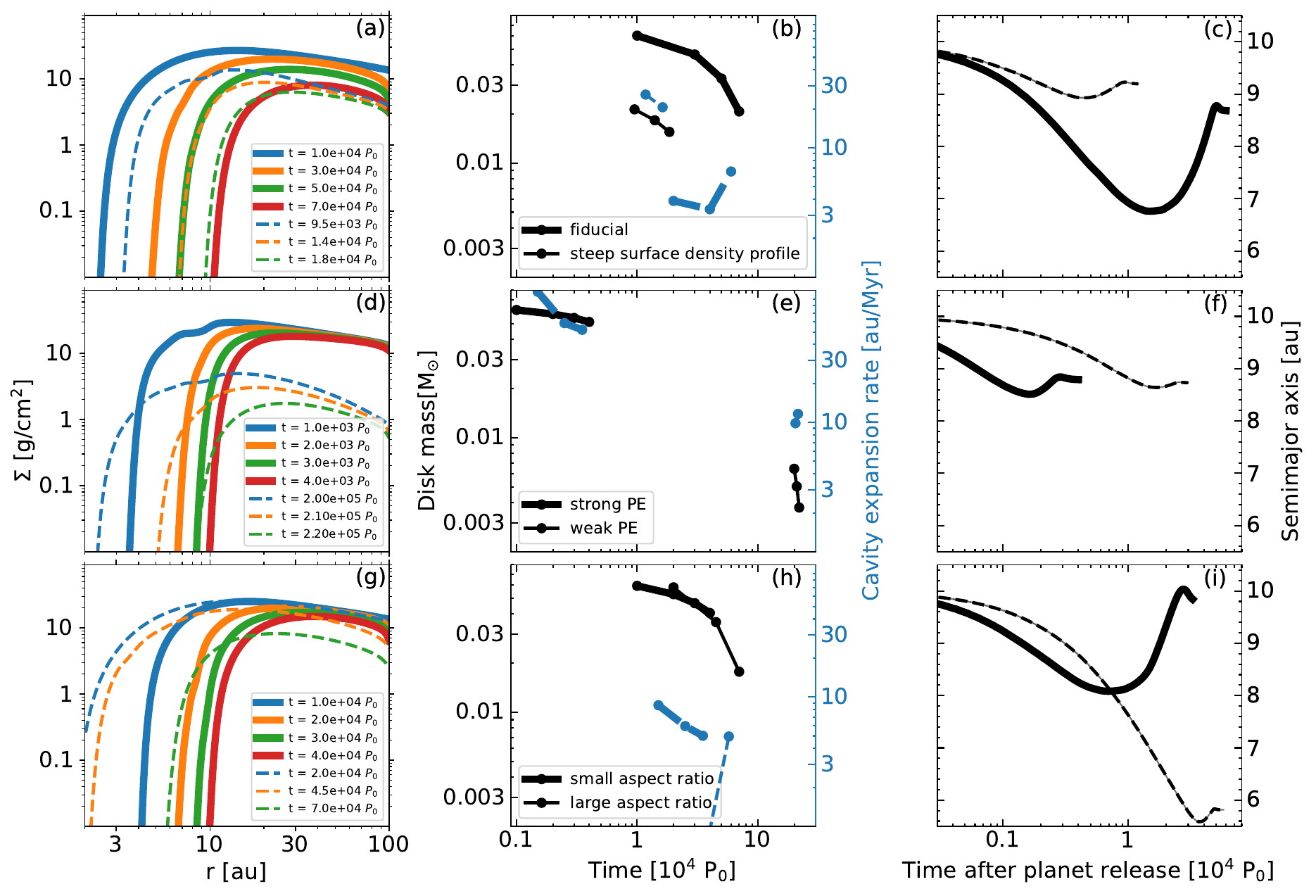}   
    \caption{Time evolution of the surface density profile of the disk (left panels), its mass (middle, black circles), the cavity's expansion rate (middle, blue circles), and the semi-major axis of a Neptune-mass planet (right). All results are shown in the PE-dominated disk phase. The planet is released at different times throughout the simulations. The three rows of panels correspond to runs with different (1) slopes of the disk's initial surface density profile, (2) PE mass-loss rates, and (3) disk aspect ratios (see \Tb{parameter}). The thick solid and thin dashed curves across the panels show results for \textit{run-fid} and \textit{run-ld} in the top row, \textit{run-ph} and \textit{run-pl} in the middle row, and for \textit{run-al} and \textit{run-ah} in the bottom row. 
    } 
    \label{fig:comparison}
\end{figure*}

 To understand how rebound migration behaves in different disk conditions,  in this section we conduct a parameter study by varying the key parameters such as the disk's surface density profile, its photo-evaporation mass loss rate and aspect ratio. Here we specifically focus on the migration of a Neptune-mass planet. The results are illustrated in \Fg{comparison} and the model setups are provided in \Tb{parameter}. 

 \subsection{Surface density profile}
 \label{sec:profile}
We carried out a simulation with an initial surface density slope of $s{=-}1$ instead of $-0.5$, keeping all other parameters the same as in the fiducial run. The results are shown in the top row of \Fg{comparison}. For $s{=}-1$ (\textit{run-dl}), the disk is initially lighter and therefore forms an inner cavity faster. At the onset of the PE-dominated disk phase, the disk mass is three times lower than in the fiducial run (see the black curves in \Fg{comparison}b). Consequently, the inner cavity expands roughly proportionally faster: the edge of the inner cavity reaches $10$ au within $2\times10^{4}~P_0$ for $s{=}-1$, and within $7\times10^{4}~P_0$ for $s{=}-0.5$ (\Fg{comparison}a). This is also seen in \Fg{comparison}b, where the blue lines show that the expansion rate of the cavity is about  $20 \rm \ au/Myr$ for $s{=}-1$, while it remains below $7 \rm ~ au/Myr$ for $s{=}-0.5$. However, the maximum density differs only by a factor $\sim$2 between both runs (about 10 vs. 20 g cm$^{-2}$). This is consistent with the fact that the inner cavity forms when the photo-evaporation rate $\dot M_{\rm PE}$  exceeds the gas inflow rate $\dot M_{\rm g}$, the latter being approximately given by $3 \pi \Sigma_{\rm PE} \nu$, with $\Sigma_{\rm PE}$ the threshold gas density when PE dominates. Thus, for disks with given $\alpha$ and $\dot M_{\rm PE}$, $\Sigma_{\rm PE}$ remains roughly the same\footnote{This also means that as long as $\Sigma_0{>} \Sigma_{\rm PE}$, varying the initial disk density has little effect on the disk evolution and planet migration in the PE-dominated phase.}. Yet, both the inward and rebound stages of migration are less pronounced in \textit{run-dl} than in \textit{run-fid} (\Fg{comparison}c). In particular, the planet only rebounds from $9$ to $9.3$ au in \textit{run-dl}, which is the consequence of the disk dispersing more rapidly and the disk density at the planet's location decreasing faster.

\subsection{Stellar photo-evaporation rate}
\label{sec:PE}
To study the influence of the photo-evaporation rate on rebound migration, we have conducted simulations with two additional $\dot M_{\rm PE}$ values: $6.4\times 10^{-9} ~ \rm M_{\odot}\,yr^{-1}$ \textit{(run-pl)} and $1.6 \times 10^{-7} ~ \rm M_{\odot}\,yr^{-1}$ \textit{(run-ph)}, using otherwise the same parameters as in our fiducial run. The results are presented in the middle row of panels in \Fg{comparison}. There are two main features regarding the disk evolution. First, the threshold surface density required to open an inner cavity is lower in the lower $\dot M_{\rm PE}$ disk, implying that the opening of the cavity takes longer than in the higher $\dot M_{\rm PE}$ disk (\Fg{comparison}d). Second, once the cavity is formed, the lower $\dot M_{\rm PE}$ disk evolves slower (\Fg{comparison}e). Rebound migration is notably much reduced in disks with high or low $\dot M_{\rm PE}$, as shown by the comparison between panels (c) and (f) in \Fg{comparison}. This can be understood as follows. If photo-evaporation is very strong, the cavity expands too quickly, the planet cannot respond to the disk evolution fast enough and thus gets left behind as the cavity sweeps out. Conversely, if photo-evaporation is very weak, the cavity expands too slowly, the timescale for gas removal at the disk edge becomes much longer than the planet's libration timescale: the corotation torque becomes weaker and outward migration cannot be sustained, making rebound migration impractical. As a result, rebound requires a moderate photo-evaporation rate to operate effectively.

Let us quantify this more. The libration timescale of the gas in the planet's horseshoe region is $t_{\rm lib} {\sim} 8 \pi r_{\rm p} /(3 \Omega r_{\rm hs})$. The typical timescale over which the disk in the horseshoe region gets depleted due to the expansion of the cavity is $t_{\rm edge} {\sim}  r_{\rm hs} / v_{\rm r}$ with $v_{\rm r}$ the expansion rate of the cavity about the planet's orbital radius. On the one hand, for the corotation torque to be sustained near its maximum value, one needs $t_{\rm edge} \lesssim t_{\rm lib}/2$, which can be recast as
\begin{equation}
\begin{aligned}
\frac{v_{\rm r}}{v_{\rm K}} \gtrsim 1.8 \times 10^{-4} \left(\frac{M_{\rm p}}{17 \ M_{\oplus}}  \right) \left( \frac{h}{0.05} \right)^{-1},
 \label{eq:case1}
 \end{aligned}
\end{equation}
where $v_{\rm K}{=} r\Omega_{\rm K}$ is the Keplerian velocity and all relevant quantities are calculated at the planet's orbital radius. On the other hand, as already said, the planet will not be able to surf the expansion of the cavity if the latter goes too fast, which sets an upper limit to $v_{\rm r}$. The latter can be estimated as the (maximum) migration rate for the planet migration to be driven by the one-sided corotation torque $\Gamma_{\rm c, 1s}$: $v_{\rm mig} {=} 2\Gamma_{\rm c, 1s} /(M_{\rm p} v_{\rm K}){=}C_{\rm hs} q_{\rm d} \sqrt{q_{\rm p}/h^3} v_{\rm K}$, where $\Gamma_{\rm c, 1s}$ is given by Eq. 11 in \cite{Liu2017a}, and $q_{\rm d}{=}\Sigma_{\rm PE} r^2/ M_{\star}$ is the ratio between the local disk mass at the onset of the PE-dominated phase and the stellar mass. The condition $v_{\rm r}{\lesssim }v_{\rm mig}$ can be recast as
 \begin{equation}
\begin{aligned}
\frac{v_{\rm r}}{v_{\rm K}} \lesssim 3.5 \times 10^{-4} \left(\frac{M_{\rm p}}{17 \ M_{\oplus}}  \right)^{1/2} \left( \frac{h}{0.05} \right)^{-3/2} \left( \frac{\Sigma_{\rm PE}}{20 \ \rm g/cm^2} \right) \left( \frac{r}{10 \rm \ au }\right)^2.
 \label{eq:case2}
 \end{aligned}
\end{equation}
Despite large uncertainties in the above timescale analysis, we can still qualitatively infer from \Eqs{case1}{case2} that rebound migration is less likely effective for (i) more massive planets, (ii) disks with a lower surface density at the onset of the PE-dominated phase, and (iii) in disks with larger aspect ratios. The latter dependence is examined in the next section.

Combing \Eqs{case1}{case2}, we estimate the critical planet mass for rebound outward migration as  
 \begin{equation}
\begin{aligned}
M_{\rm rebound} \simeq 64     \left( \frac{h}{0.05} \right)^{-1} \left( \frac{\Sigma_{\rm PE}}{20 \ \rm g/cm^2} \right)^2 \left( \frac{r}{10 \rm \ au }\right)^4  M_{\oplus}.
 \label{eq:critp}
 \end{aligned}
\end{equation}
\Fg{pmass} illustrates how this planet mass varies with radial distance and aspect ratio at given $\alpha$ and $\Sigma_{\rm PE}$. The black solid line denotes $M_{\rm  rebound}$ while the colored dashed lines represent the gap opening mass $M_{\rm gap}$ from \cite{Crida2006}, \cite{Fung2014} and \cite{Kanagawa2015}, respectively. The intersection region below both the colored and solid lines is expected to be optimal for rebound migration.

 \begin{figure*}
    \centering
        \includegraphics[width=12cm]{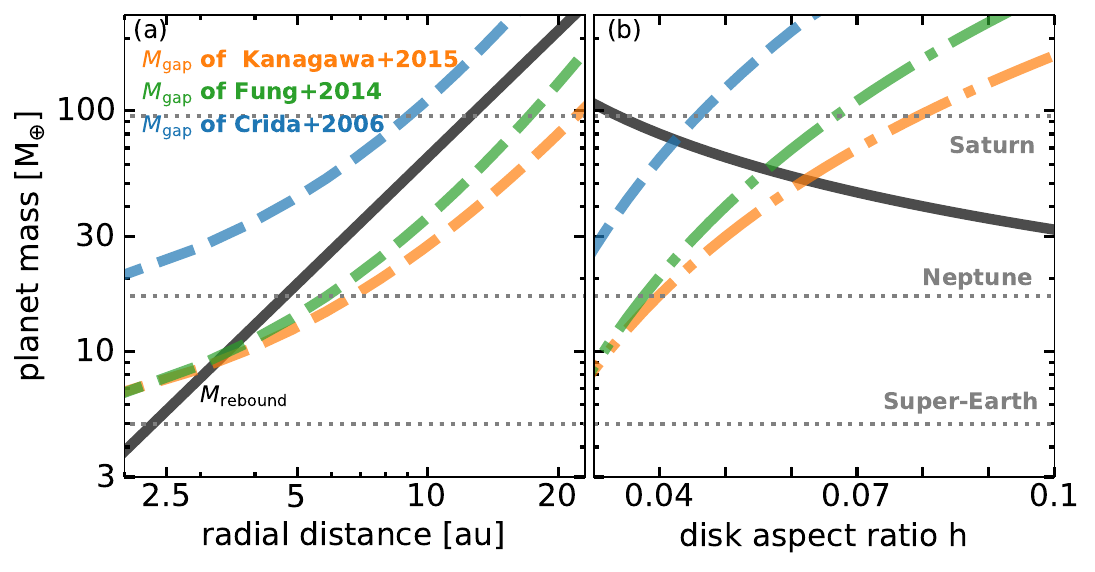}
    \caption{Planet mass vs radial distance (panel a) and disk aspect ratio (panel b). The solid line represent the critical planet mass for rebound migration based on our \eq{critp}. The blue, green and orange dashed lines give the fitting formulas of \cite{Crida2006}, \cite{Fung2014} and  \cite{Kanagawa2015}, where the first one and the latter two define $M_{\rm gap}$ as the mass corresponding to $10\%$ and $50\%$ of the gas density at the center gap compared to the unperturbed values. The other parameters are $\alpha{=}10^{-3}$, $\Sigma_{\rm PE}{=}20 \times  (r/10 \ \rm au)^{-1} \rm \ g/cm^{-2}$ at $h{=}0.05$ in panel a and $r{=}10 $ au in panel b, respectively.     
    } 
    \label{fig:pmass}
\end{figure*}

\subsection{Disk aspect ratio}
\label{sec:aspectratio}

We have also tested the influence of the disk's aspect ratio by varying $h_0$ to $0.04$ (\textit{run-al}) and $0.07$ (\textit{run-ah}), again with all other parameters being those of the fiducial run. Results are shown in the bottom row of \Fg{comparison}. In both additional simulations, the inner hole forms at a similar threshold (peak) surface density. However, the lower the aspect ratio, the earlier the cavity forms, and the faster the disk subsequently evolves. This can be explained by considering photo-evaporation as a competition between gravitational energy, which binds the gas, and thermal energy, which drives the gas outflow. Hence, disks with lower aspect ratios and thus lower temperatures experience faster evaporation. Besides, the magnitude of the disk torque on the planet increases with decreasing aspect ratio. Combining these two effects, we find in \Fg{comparison} that the relative magnitude of rebound migration is similar in \textit{run-al} to that in the fiducial run, while it is much lower in simulation \textit{run-ah}. In the latter case, the planet undergoes outward migration only at a few percent level relative to its initial semi-major axis.

\section{Summary and Conclusion}
\label{sec:conclusion}

In this study, we have presented the first 2D hydrodynamical simulations of disk-planet interactions showing that outward "rebound" migration may occur indeed near the edge of an expanding cavity in a protoplanetary disk due to photo-evaporation from the central star. The combined effects of turbulent viscous accretion and stellar photo-evaporation cause the formation of an inner hole at orbital distances of a few au. Subsequently, the disk outside the cavity experiences rapid mass loss due to the evaporative wind generated by high-energy stellar radiation. As the gas disk disperses, the inner edge of the disk expands inside-out. Importantly, a strong positive density gradient builds up along the edge of the cavity, which can largely impact planet migration.

To investigate planet migration under these circumstances, we have carried out hydrodynamical simulations with the $2$D code \texttt{Dusty FARGO-ADSG}, in which we have implemented the effects of stellar X-ray photo-evaporation. We have examined the migration behaviour of a single planet with four characteristic masses: super-Earth, Neptune, Saturn, and Jupiter. We have analysed the influence of various disk parameters, such as the disk's surface density profile, the stellar photo-evaporation mass-loss rate, and the disk's aspect ratio.

For the super-Earth and Neptune-mass planets, a large positive corotation torque due to the strong density gradient along the cavity edge is able to sustain substantial outward "rebound" migration. Our 2D hydrodynamical simulations thus confirm the possibility of rebound migration, which was originally proposed and analytically calculated by \cite{Liu2017a}. The amplitude of rebound migration is found to increase with planet mass, with lower disk aspect ratio, and requires a moderate photo-evaporation rate. For Saturn-mass planets, however, no rebound migration is observed in our simulations, which we attribute to partial gap depletion lowering the magnitude of the corotation torque. The Jupiter-mass planet can migrate outward along the expanding cavity, which is largely due to the outer disk becoming eccentric. Future work is needed to further investigate rebound migration, in particular for multiple planet systems in resonant chains with different masses and under various disk conditions.

\begin{acknowledgements} 
This work is supported by the National Key R\&D Program of China (2024YFA1611803). BL acknowledges support from the National Natural Science Foundation of China (Nos. 12222303 and 12173035) and the start-up grant of the Bairen program from Zhejiang University. The simulations and analysis presented in this article were carried out on the SilkRiver Supercomputer and the Qilin computing cluster of Zhejiang University, and on the CALMIP Supercomputing Centre of the University of Toulouse.
Y.-P. L. is supported in part by the Natural Science Foundation of China (Grants No. 12373070 and No. 12192223), and the Natural Science Foundation of Shanghai (Grant No. 23ZR1473700).
\end{acknowledgements}

\bibliographystyle{aa}
\bibliography{reference.bib}
\end{document}